# SynthBrainGrow: Synthetic Diffusion Brain Aging for Longitudinal MRI Data Generation in Young People

Anna Zapaishchykova[1,2,3*], Benjamin H. Kann[1,2,3*],
Divyanshu Tak[1,2,3], Zezhong Ye[1,2,3], Daphne A. Haas-Kogan[1,2,4],
Hugo J.W.L. Aerts[1,2,3]

[1] Artificial Intelligence in Medicine (AIM) Program, Mass General Brigham, Harvard Medical School, Boston, MA, United States
[2] Department of Radiation Oncology, Dana-Farber Cancer Institute and Brigham and Women's Hospital, Harvard Medical School, Boston, MA, United States
[3] Radiology and Nuclear Medicine, CARIM & GROW, Maastricht University, Maastricht, the Netherlands
[4] Boston Children's Hospital, Boston, MA, United States

```
azapaishchykova@bwh.harvard.edu,
benjamin_kann@dfci.harvard.edu
```

**Abstract.** Synthetic longitudinal brain MRI simulates brain aging and would enable more efficient research on neurodevelopmental and neurodegenerative conditions. Synthetically generated, age-adjusted brain images could serve as valuable alternatives to costly longitudinal imaging acquisitions, serve as internal controls for studies looking at the effects of environmental or therapeutic modifiers on brain development, and allow data augmentation for diverse populations. In this paper, we present a diffusion-based approach called SynthBrainGrow for synthetic brain aging with a two-year step. To validate the feasibility of using synthetically-generated data on downstream tasks, we compared structural volumetrics of two-year-aged brains against synthetically-aged brain MRI. Results show that SynthBrainGrow can accurately capture substructure volumetrics and simulate structural changes such as ventricle enlargement and cortical thinning. Our approach provides a novel way to generate longitudinal brain datasets from cross-sectional data to enable augmented training and benchmarking of computational tools for analyzing lifespan trajectories. This work signifies an important advance in generative modeling to synthesize realistic longitudinal data with limited lifelong MRI scans. The code is available at XXX.

## 1 Introduction

Brain aging research relies heavily on magnetic resonance imaging (MRI) to track longitudinal changes in brain structure and function [1], [2]. Modeling long-term trajectories of different volumetric structures is critical for understanding healthy development, neurodegenerative disorders [3] and the effect of interventions on brain development [4]. However, such lifelong longitudinal MRI data remains scarce.

---

*These authors contributed equally to this manuscript

Recent advances in generative modeling provide new opportunities to synthesize pseudo-longitudinal MRI data simulating brain aging effects. Denoising diffusion probabilistic models (DDPMs) have shown early promise in the synthetic MRI generation [5]. Comparing synthetic MRIs that represent healthy brain aging to actual patient clinical scans could reveal neurodevelopmental diseases, abnormalities due to environmental or clinical interventions such as psychiatric medications or radiotherapy, patterns of atrophy, and other biomarkers associated with neurodegenerative diseases [4].

In this work, we propose a conditional DDPM that takes 3-dimensional (3D) brain T1w MRIs as input and generates synthetic images simulating the subject brain maturation two years into the future. Our model was trained on paired scans from individuals two years apart in the demographically diverse sample from the Adolescent Brain Cognitive Development (ABCD) study [6]. By learning transformations from the first to the second scan showing natural aging effects and utilizing the input volume as a conditional guidance, our model can generalize to new input scans and output images depicting simulated aging.

This approach could generate pseudo-longitudinal data, augmenting existing MRI studies and databases. In addition, our synthesized aged brains could provide controllable test cases for evaluating computational analysis tools focused on volumetric changes over time. Visualizations of normal versus abnormal aging trajectories from our model may provide clinical decision support.

**Related Work.** Physical pubertal maturation was previously reported to be associated with brain development beyond chronological age [7]. ABCD study spans across 21 research sites across the United States, has been used to investigate various aspects of adolescent health and behavior, such as sociocultural influences on alcohol expectancies, associations resting-state functional brain connectivity, and childhood anhedonia [8], [9].

Recently, DDPMs have gained much attention due to their superior performance in image generation [10]. DDPM was further improved by changing the loss objective, making architecture improvements, and using classifier guidance during sampling, improving the output image quality [11]. In medical imaging, DDPM has shown success in various tasks, such as segmentation [12], under-sampled medical image reconstruction [13], estimating brain age from routine MRI [14], and contrast harmonization [15]. Synthetic MRI has been explored to evaluate changes in relaxation values in different brain regions and construct brain age prediction models [16]. While some new research applies diffusion models to tasks such as medical image generation [5], to our knowledge, there is no preliminary work on synthetic brain aging in children through young adulthood.

Recently, one approach was proposed by Fu et al. that explores the generation of synthetic brain aging images by diffeomorphic registration, enabling the augmentation of 3D MRI scans for healthy brain aging for adult subjects [17]. However, since their method relies on diffeomorphic registration, it requires two images, unlike generative approaches. In contrast to their work, we focus on the younger subjects ranging from 8-16 years old from the ABCD study with a two-year scan interval and DDPM as a backbone model, which allows us to synthetically age brain MRI using only one baseline image.

**Contribution.** We propose the first diffusion model for the synthetic aging of subject-specific brain MRIs and the first model of any kind for synthetic aging in young people. Our model simulates two years of anatomically-plausible brain maturation based on paired scans showing real aging effects. We demonstrate the utility of our synthesized pseudo-longitudinal data by analyzing age-related substructural volumetrics and volumetric changes.

## 2 Method

An overview of the workflow for an image of the ABCD dataset is shown in **Fig. 1**.

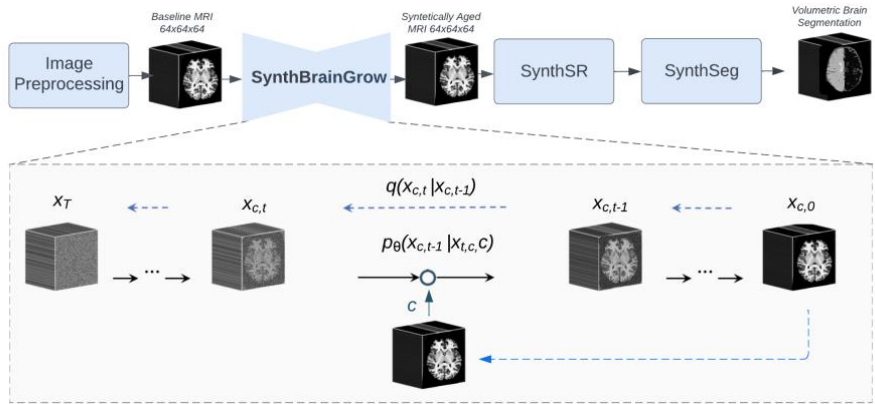

**Fig. 1. Top panel:** Method overview. Step 1: MRI preprocessing, pairwise co-registration, intensity normalization, and rescaling to 64x64x64. Step 2: Diffusion probabilistic model SynthBrainGrow for synthetic brain 2-year aging. Step 3: Image upscaling using SynthSR. Step 4: Brain tissue segmentation using SynthSeg. For more details on each step, please refer to the "3.1 Experimental Setup & Dataset" section. **Bottom panel:** The training and sampling procedure of our method. In every step $t$, the anatomical information is induced by concatenating the baseline brain MR images $b$ to the noisy aged brain $x_{c,t}$

Our model was trained on paired 3D T1w MRI scans of the same subjects scanned two years apart. The first scan provides the input for the baseline of healthy brain, while the second scan provides the ground truth for the image after two years of aging. By training the diffusion model on these input-output pairs, the model learns to take a healthy brain as input and output a version that has simulated two years of aging. We follow the idea and implementation proposed by Wolleb et al. [12] and Dorjsembe et al. [18]. Like DDPMs, our aging synthesis approach relies on a forward diffusion process that adds Gaussian noise to brain MRI scans from young healthy individuals, followed by a reverse generative process that denoises the images. However, we incorporated the anatomical guidance from the baseline scan during diffusion. Specifically, at each time step t, our model takes as input a noisy aged brain image $x_t$ along with a corresponding input baseline brain scan $c$. We concatenate these along the channel dimension to produce an augmented input:

$$X := x_c \oplus c \tag{1}$$

This concatenated volume provides essential anatomical cues to guide the denoising diffusion process. The forward diffusion process that corrupts the baseline scan $x_0$ over T steps is defined the same as DDPM:

$$x_{c,t} = \sqrt{\bar{a}_t} x_0 + \sqrt{1-\bar{a}_t} e, \quad with\ e \sim N(0, I) \tag{2}$$

The reverse generative modeling process relies on our conditional diffusion model $p_\theta(x_{t-1}|\sim x_t)$. At each timestep, the model takes as input $\sim x_t$ and outputs the denoised $x_{t-1}$ used for generation after T steps:

$$x_{t-1} \sim p_\theta(x_{c,t-1}|\sim x_{t,c} c) \tag{3}$$

Through exposure to anatomical conditional guidance during diffusion, we hypothesize $p_\theta$ will learn mappings to synthesize aged scans. The loss objectives and model hyperparameters are specified in the appendix of [11]. Due to the stochastic nature of the DDPM, aging twice for the same brain MR image $c$ does not result in the same output.

## 3   Experiments and Results

### 3.1   Experimental Setup & Dataset

We evaluated our method on the ABCD dataset (Data Release 5.1). The ABCD Study® operates as a consortium, comprising 21 data collection sites across the continental US to sample in an epidemiologically-informed and inclusive way [6]. We performed a pairwise registration using the Elastix [19] package for each patient 3D T1w MRI scans pair, followed by a skull stripping step using HD-BET [20]. The image intensity was then normalized with brain mask as guidance. Addtionally, the image was downsampled to 3×3×2.5 mm$^3$ in voxel size and the resulting volume was cropped to the size of 64×64×64 mm$^3$. To overcome the memory size constraints and save computational time during model training, we pre-computed all the preprocessing steps prior to the deep learning training.

The total number of 3D T1w MRI pairs is 9324, originating from 7843 patients aged 8-16 years (53% Male). We performed the random 70/15/15 train/validation/test split, which results in 6526/1399/1399 MRI scan pairs. We chose a linear noise schedule for T=1000 steps. The U-Net was trained with the loss objectives given in the study by Nichol et al. [11] using the MONAI framework v1.4 with a learning rate of $10^{-4}$ using Adam optimizer and a batch size of 1. We trained the model for 4,000 epochs on 1x Nvidia A6000 with a validation evaluation step for every 100 epochs, which took around one day per 100 epochs.

For the MRI postprocessing, we upsampled the image ×2 using spline interpolation, resample voxel size back to 1×1×1 mm$^3$, and increased image resolution using FreeSurfer v.7.4.1 SynthSR v2.0 [21]. To segment brain structures, we used FreeSurfer v.7.4.1 SynthSeg v1.0 [22] (see **Fig. 2**. for an example of synthetically aged brain MRI). We discarded the testing cases with anatomically-implausible ground truth segmentation, which is lower than 30 WMV and lower than 1 mm$^3$/10,000 sGMV units. All implementation details can be found in the study git repository XX.

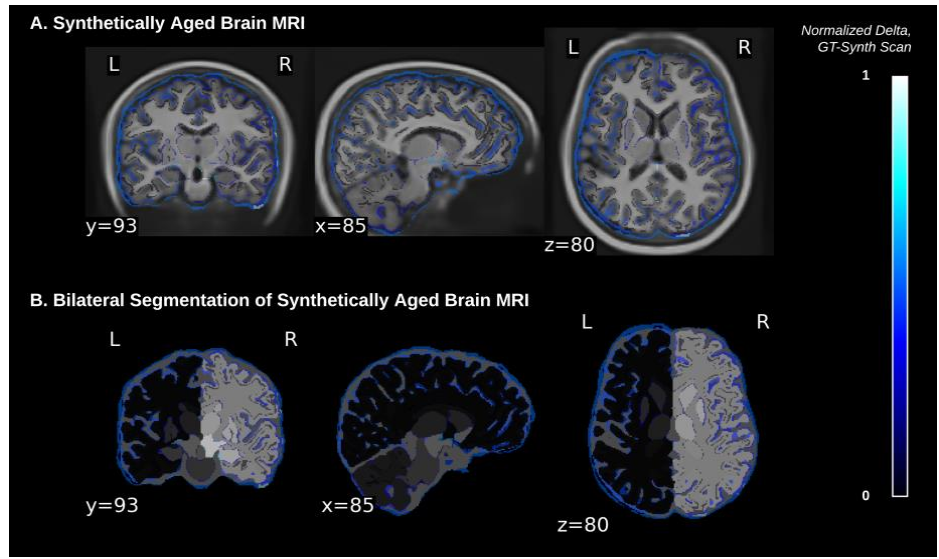

**Fig. 2.** A. An example of synthetically-aged brain MRI in axial, sagittal, and coronal view cuts (z=80, x=85, y=93) with an overlaid heatmap (blue) of the normalized delta, which was calculated as the difference between a ground truth scan and a synthetically-aged scan. A lighter color indicates more difference. **B.** SynthSeg bilateral segmentation mask of synthetically-aged scan with an overlay heatmap (blue) of the normalized delta.

### 3.2   Quantitative Image Quality

The evaluation of synthetic medical image quality requires robust metrics to ensure accuracy and reliability. The use of structural similarity indices such as the structural similarity index measure (SSIM) for evaluating synthetic medical images has come under recent scrutiny, as it may not effectively capture perceptual quality or clinical usefulness in synthesized radiology scans [23]. This limitation seems especially relevant for synthetic brain MRIs modeling neurodevelopment, where clinical value is derived from quantitative biomarkers like volumetrics [24]. Similarly, SSIM does not reflect image quality well, suggesting its inadequacy in evaluating image quality in certain contexts [25], [26]. To assess the performance of SynthBrainGrow, we evaluated the substructural volumetric similarity between synthetic and real patient scans. We compared total gray matter (GMV), white matter (WMV), subcortical gray matter (sGMV), and ventricular (VV) volumes by calculating Pearson correlation and mean absolute error (MAE) between our aging model-predicted outputs and anatomically corresponding ground truth validation scans (**Table 1**).

**Table 1.** Volumetric structural comparison between ground truth intra-subject two-year aged brains and synthetically-generated ones (N=1399). GMV: Gray matter GMV; WMV: white matter; sGMV: subcortical GM; VV: ventricular volume; MAE: mean absolute error; Delta,%: the difference between synthetically generated one and ground truth intra-subject two-year-aged brain, normalized by the ground truth.

| Structural Volumetrics | Pearson R | MAE, $mm^3/10,000$ | Delta, % |
|---|---|---|---|
| WMV | 0.89 | 0.95 | 0.1 |
| GMV | 0.74 | 4.5 | 0.08 |
| sGMV | 0.45 | 0.55 | 0.07 |
| VV | 0.83 | 0.18 | 0.04 |

Strong volumetric correlations were observed in WMV, GMV and VV with Pearson R values (p<0.05) from 0.74 (GMV) to 0.89 (WMV), demonstrating that the SynthBrainGrow accurately generates realistic patterns of the aging process (**Fig. 3).** A moderate correlation with Person R 0.45(p<0.05) was observed for sGMV. Mean absolute volume errors between synthetic and real patient scans were in the range of 0.2 mm3/10,000 (VV) and 4.8 mm3/10,000 (GMV), indicating good volumetric validity.

### 3.3    Uncertainty Maps as an Explainability Surrogate

By utilizing the inherited property (stochastic sampling process) of the DDPMs, we can generate a distribution of aged 3D MRI scans. This property allows us to compute pixel-wise uncertainty and allows an implicit ensemble to show patient-specific regions of interest contributing to brain maturation and, therefore, help to bridge the "black-box" explainability gap in DL. In **Fig. 4.**, we visualize a variance map by predicting the aged brain of one subject ten times. The regions that are most affected by the diffusion aging model appear to correspond with areas where structural changes related to aging occur, such as enlargement of the ventricles and cortical thinning. The clinical utility of attention maps remains to be evaluated, but they may provide an interpretable output and could potentially be used for uncertainty quantification when used as a visual guide.

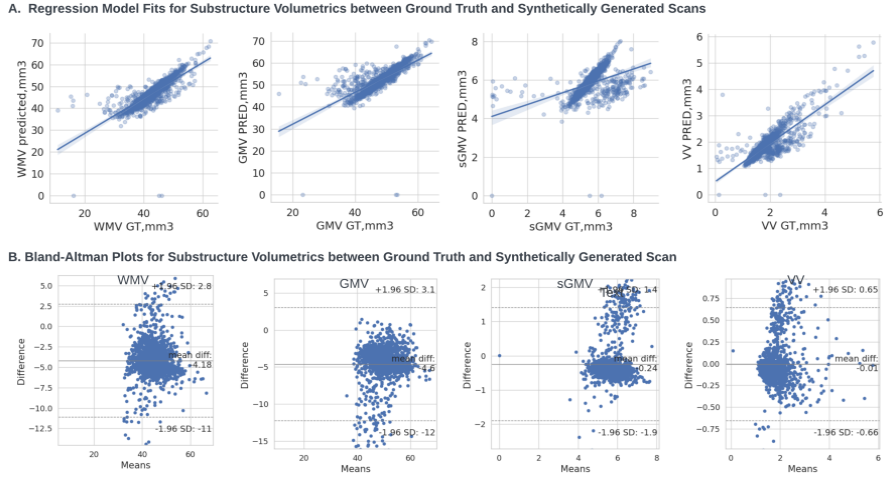

**Fig. 3. A.** Scatterplots with regression model fit lines comparison of ground truth (GT) versus synthetically-aged (prediction) scan for bilateral WMV, GMV, sGMV, and VV volume (N=1399). Axes are scaled in units of 10,000 mm$^3$. **B.** Bland-Altman plots for substructure volumetrics agreement between GT vs. prediction for bilateral WMV, GMV, sGMV, and VV. GMV: Gray matter GMV; WMV: white matter; sGMV: subcortical GM; VV: ventricular volume.

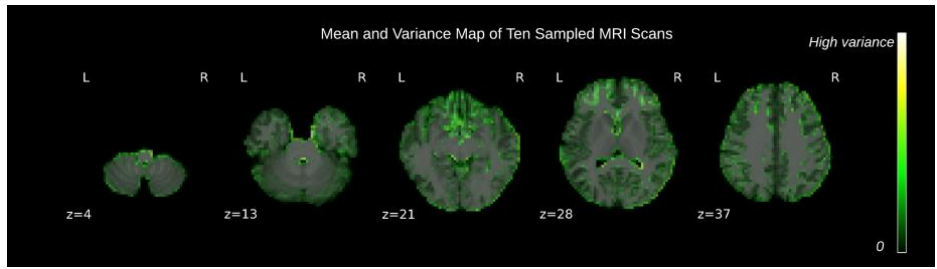

**Fig. 4.** Combined uncertainty mean maps with variance heatmap overlay for ten sampled MRI T1w brains for a single subject, five axial slice views (z- axial slice number). A lighter color means higher variance.

### 3.4 Limitations and Future Directions

Our model was trained on a relatively narrow age range and sample from one study representative of the population within United States. Testing performance is needed when extrapolating beyond the training data to younger or older ages. Real longitudinal within-person trajectories may show more variability and nonlinearity than model approximation. Incorporating diverse scans from multi-site datasets spanning different demographics, health statuses, and neurodegenerative conditions might reveal where synthesis quality drops and additional training is required.

Mapping synthetic scans back to brain age versus chronological age biomarkers may offer a universal framework for validation. Ideal outputs would mirror consistent but variable patterns of within-person maturation and decline in large-scale studies. This could indicate utility for personalized prediction of neurocognitive trajectories. Extending to longitudinal training and evaluating scan trajectories against real neuropsychological, molecular, and clinical aging biomarkers is an exciting future direction.

Additionally, we will consider sampling with the DDIM approach to speed up the sampling process in future work.

## 4    Conclusion

We developed a generative model approach SynthBrainGrow for synthetic 2-year brain maturation in MRI T1w. Using a stochastic sampling process, our method enables the generation of different MRIs for the same input brain MR image without training a new model. Moreover, the model yields uncertainty maps by computing the variance to measure clinical interpretability. Our results suggest that synthetically-aged brain MRI with diffusion accurately captures substructure volumetric trends and could be used as a control for studies investigating modifying factors of brain development. For future work, we plan to explore the aging process in subjects with brain abnormalities and expand our dataset to a broader age range. The next priority is assessing generalizability and clinical relevance by evaluating performance on diverse unseen target groups and prediction intervals. For now, this model framework can be utilized and fine-tuned by the research community to generate short-interval brain aging effects in various scenarios.

**Acknowledgments.**

**Disclosure of Interests.**